\documentstyle[12pt]{article}
\begin{document}
\baselineskip 23pt
\begin{center}
\vspace*{3cm}
{\Large \bf Pinning of a solid--liquid--vapour \\ interface by stripes of
obstacles.}\\
\vskip1cm
{\bf J\o rgen Vitting Andersen} \\[3mm]
Laboratoire de Physique de la Mati\`{e}re Condens\'{e}e, \\
Universit\'{e}\ de Nice-Sophia Antipolis,  \\
Parc Valrose, 06108 Nice Cedex 2, France\\[2mm]
{\bf Yves Br\'{e}chet}  \\[3mm]
Laboratoire de Thermodynamique et Physico-Chimie M\'{e}tallurgique, \\
Institut National Polytechnique de Grenoble, BP 25, \\
Domaine Universitaire de Grenoble,
38402 St. Martin d'H\'{e}res,
France \\[1cm]
\end{center}

\begin{center} {\bf Abstract}\end{center}
We use a macroscopic Hamiltonian approach to study the pinning of
a solid--liquid--vapour contact line on an array of equidistant stripes of
obstacles perpendicular to the liquid. We propose an estimate
of the density of pinning stripes for which collective pinning of
the contact line happens. This estimate is shown to be in
good agreement with Langevin equation simulation of the macroscopic
Hamiltonian. Finally we introduce a 2--dimensional mean field theory
which for small strength of the pinning stripes and for small capillary
length gives an excellent description of the averaged height of the contact
line.

\noindent PACS numbers: 68.10.-m, 68.45.Gd,68.45Ws.

\newpage

\section{\bf Introduction}
The spreading (wetting) of a liquid on a solid is important in a
widespread field of practical applications such as lubrication, the
efficiency of detergents, oil recovery in a porous medium and the
stability of paint coatings\cite{deGennes}. The motion of the
interface  is often  extremely sensitive to impurities and roughness,
which tend to pin (stick) the interface.
Different situations arises for the motion of the boundary line between a
solid, liquid and vapour (called the triple line or the contact line) depending
on the
heterogeneousity of the solid  and
depending on whether the liquid completely wets the
solid surface\cite{completewetting} or incompletely wets the solid[3-13].
Due to the presence of a microscopic precursor film that advances ahead of
the macroscopic liquid, the case of complete wetting is noticeable insensitive
to
the heterogeneousity of the solid.
In the incomplete wetting case obstacles tend to pin the contact line
which makes the statics and dynamics of the contact line highly sensitive
to the specific form of the heterogeneousity of the solid.
Similar
kinds of problems are also encountered in other situations where
an elastic body is pinned by a random potential\cite{pinningrefs} such as flux
pinning
of type-II high $T_c$ superconductors\cite{Larkin},
pinning of charge density waves\cite{Sneddon}, pinning
of magnetic domain walls\cite{Grinstein}
 and dislocation pinning\cite{Kocks}.

We have proposed
a macroscopic Hamiltonian approach to study the
pinning of a solid--liquid--vapour
interface when a solid is pulled vertically out of a liquid (called the
immersion
geometry). We investigate the simple case of the statics of an array of
equidistant
stripes of obstacles, and we study how the averaged height of the triple line,
depends
on the capillary length and the density and strength of the stripes.
We hope by studying this simple case to understand some of the important
physics that is involved in the more complex case of completely random pinning
sites.

\section{\bf A macroscopic Hamiltonian approach to the pinning of
solid--liquid--vapour
interfaces}

The geometry of the problem is chosen so that the liquid--vapour interface is
in the
$x$-$y$-direction and the solid is pulled vertically out of the liquid in the
$z$-direction.
The origin of the $z$-axis is taken to coincide with the liquid-vapour
interface.
The local surface energies for the liquid--gas interface, $\gamma_{LG}$, the
solid--liquid
interface, $\gamma_{SL}(y,z)$ and the solid--gas interface $\gamma_{SG}(y,z)$
determine
a local force balance expressed by Young's\cite{Young} relation:
\begin{eqnarray}
\gamma_{LS} (y,z) + \gamma_{LG} \cos \theta(y,z) & =& \gamma_{SG} (y,z)
\label{young}
\end{eqnarray}
$\theta(y,z)$ is the local macroscopic contact angle between the solid and the
liquid,

  The Hamiltonian reads
\begin{eqnarray}
{\cal H}& =& \int_{0}^{L}\int_{0}^{M} dx dy {1 \over 2} \rho g h^2(x,y)
+ \gamma_{LG} \int_{0}^{L}\int_{0}^{M} dx dy \sqrt{1 + (\vec{\nabla} h(x,y))^2}
\nonumber \\
& & + \int_{0}^{M}\int_{0}^{h(0,y)} dy dz \gamma_{SL} (y,z)
+ \int_{0}^{M}\int_{h(0,y)}^{N} dy dz \gamma_{SG} (y,z) ,
\label{hamiltonian}
\end{eqnarray}
where $\rho$ is the mass density of the liquid (assumed constant) and $g$ the
gravity
constant.
The first term in this expression is the gravity potential energy, the second
term
describes the surface energy between the liquid and the gas, and the last two
terms
account for the surface energies due to contact between the solid wall
and the liquid, respectively, gas phase\cite{note2}.

If one considers how the Hamiltonian changes by a small change
in $\delta h(x,y)$,
the three parameters $\gamma_{SL} (y,z),
\gamma_{SG} (y,z)$ and $\gamma_{LG}$ that enters the Hamiltonian,
can be expressed in terms of just two variables, namely
the liquid--gas surface energy, $\gamma_{LG}$, and the macroscopic contact
angle
$\theta (y,h(x,y))$ using Young's relation
Eq.~(\ref{young}):
\begin{eqnarray}
{\delta {\cal H} \over \delta h(x,y)}& =& \rho g h(x,y)
- \gamma_{LG} \nabla^2 h(x,y) (1 + (\vec{\nabla} h(x,y))^2)^{-{1 \over 2}}
\nonumber \\
& & + \gamma_{LG} \nabla^2 h(x,y) (1 + (\vec{\nabla} h(x,y))^2)^{-{3 \over 2}}
(\vec{\nabla} h(x,y))^2
-\gamma_{LG} \cos (\theta (y,h(x,y))) \delta (x)
\label{delham}
\end{eqnarray}
The sole quantity that has the dimensions of a length is called the capillary
length and
is defined as:
\begin{eqnarray}
\kappa & \equiv & \sqrt{ 2 \gamma_{LG} \over \rho g}.
\label{kappa}
\end{eqnarray}
On the length scales for which the Hamiltonian Eq.~(\ref{hamiltonian}) is
supposed
to be valid, one can safely ignore fluctuations due to the temperature, so the
problem
of finding the configuration $h(x,y)$ that minimizes ${\cal H}$ is a zero
temperature problem.

One way to obtain the equilibrium configuration $h(x,y)$ described by the
Hamiltonian
Eq.~(\ref{hamiltonian}) is to
perform simulated annealing using Monte Carlo
simulation.
Notice that since only the {\em change} of
the total energy is needed in a Monte Carlo update, the equilibrium state is
completely specified by the three variables $\rho g, \theta(y,z)$ and
$\gamma_{LG}$.
In Ref.~\cite{ego} simulated annealing was carried out for the case
$\theta(y,z)
= constant$, and for the case of equidistant stripes of obstacles
(with contact angle
$\theta'$) in the $z$-direction.
The case $\theta(y,z)
= constant$ served as a check of the validity of Eq.~(\ref{hamiltonian}) since
the profiles $h(x,y)$ can be directly compared to various analytical
results\cite{landau}. Initial findings for the case of stripes of obstacles
indicated that the averaged height of the triple line, $<h(0,y)>$, was
linear in the density of pinning sites, $c$,
for small values of $c$ and with a crossover to nonlinear
behavior for $c \rightarrow 1$. The density for which the crossover
happened, $c^*$, was an increasing function of ${1 \over \kappa}$. This is
to be expected, since for small $c$ a given pinning stripe does not feel the
presence of the other pinning stripes, and $<h(0,y)>$ can be obtained
as a simple superposition over all the pinning stripes of the profiles
of individual stripes that pin the liquid--vapour interface.
On the other hand when $c$ becomes larger the pinning stripes mutually
(collectively) lift
the liquid, and the resulting profile $h(0,y)$ can not be obtained as
a simple superposition over the pinning stripes.
Therefore one
expects
that collective pinning sets in once
the averaged distance between pinning stripes, $d$, becomes of the order of
the capillary length $\kappa$, giving $c^* \sim 1/d \sim 1/\kappa$. We will
in this paper show that another length scale enters the problem, so that the
before mentioned argument has to be modified.

It turns out that simulated annealing has the disadvantage of requirering
large amounts of computing time, since one needs to perform a slow annealing
sequence $T(t)$\cite{note1} in order to bypass metastable configurations.
Furthermore the optimal sequence $T(t)$ depends on the parameters $\theta$,
$\theta'$, $c$, $\kappa$, the lattice constant $a$,
and one should in principle determine the optimal sequence $T(t)$
and the optimal step size $\Delta h$  for each
set of parameters values used in a simulation.
In this paper we have instead used Langevin equation simulations of
Eq.~(\ref{delham}) since it was found to be computational more efficient than
the Monte Carlo simulations. That is, we have numerically solved the equation
\begin{eqnarray}
{\delta h(x,y) \over \delta t} & =& -\Gamma {\delta {\cal F} \over \delta
h(x,y)} .
\label{langevin}
\end{eqnarray}
$\Gamma$ is a mobility constant, $t$ is the time which we let go to infinity in
order
to find the equilibrium state, and ${\cal F}$ is the free energy of the system
which we will assume ${\cal F} \approx {\cal H}$, since we neglect
fluctuations due to the temperature.
In order to take proper account of the boundary
condition for the contact of the liquid with the solid plate one has that
\begin{eqnarray}
{\partial h(x,y) \over \partial x}\mid_{x \rightarrow 0} = \cot (\theta
(y,h(0,y))).
\label{bc}
\end{eqnarray}
This condition has to be introduced via a Langrange multiplyer as an
extra term to the Hamiltonian Eq.~(\ref{hamiltonian}), or equivalently one
leaves out the contact terms in the energy Eq.~(\ref{hamiltonian}) (which
amounts
to leave out the term $-\gamma_{LG} \cos (\theta (y,h(x,y))) \delta (x)$ in
Eq.~(\ref{delham})) and expresses them instead via Eq.~(\ref{bc})).

When discretizing Eq.~(\ref{langevin}) the functional form of theta for a
realization
of stripes of obstacles
takes the form
\begin{eqnarray}
\theta (y,z) & = & \delta_{{\rm modulus}(y,i),0} \theta'
+ (1-\delta_{{\rm modulus}(y,i),0}) \theta ,
\label{modtheta}
\end{eqnarray}
with $\theta'$ the value for the contact angle on the stripes of obstacles,
$\theta$
the value of the contact angle between the stripes of obstacles,
and $1 \leq i \leq L/a$ an integer
determining the density ($0 \leq c \leq 1$) of pinning stripes. One notice that
due
to translational invariance in the $z$-direction, there is no pinning force
acting
when the liquid is pulled vertically out of the liquid, whereas a rotation of
the solid
about an axis in the $x$-$y$-plane or a translation of the solid in the
$y$-direction
produces a pinning force.

\section{\bf Results}
\begin{figure}
\begin{center}
\caption{The averaged height of the triple line $<h>$ versus the density $c$
for
different values of $\kappa$.
$\kappa =4 (\Box)$, $\kappa =2 (+)$ and $\kappa =1 (\Diamond)$ respectively.
$\theta = {\pi \over 2}, \theta' = 1.5.$}
\end{center}
\end{figure}
\begin{figure}
\begin{center}
\caption{The averaged height of the triple line $<h>$ versus the density $c$
for
different values of $\kappa$.
$\kappa =1 (\Diamond)$, $\kappa =0.5 (+)$,
$\kappa =0.25 (\Box)$ and $\kappa =0.125 (\times)$ respectively.
$\theta = {\pi \over 2}, \theta' = 0.5.$
{}From Eq.~(\ref{c*curvature}) one finds:
$c_{\rm curvature}(\kappa =1.0)=0.15,
c_{\rm curvature}(\kappa =0.5)=0.30,
c_{\rm curvature}(\kappa =0.25)=0.60$,  and
$c_{\rm curvature}(\kappa =0.125)>1$.}
\end{center}
\end{figure}

In Fig.~1 is shown
the averaged height of the triple line $<h>$
versus the density $c$ for
different values of $\kappa$ from a
numerical calculation of Eq.~(\ref{langevin},\ref{bc},\ref{modtheta}).
The discretization parameter
was chosen
$a=0.04$ (the same for all the results represented in this paper), a typical
time step $\Delta
t \approx 10^{-4}-10^{-2}$  depending on the values of
$\kappa , \theta , \theta' , c$; the lattice sizes were chosen between
$L/a \times M/a = 100 \times 100$ and $L/a \times M/a = 100 \times 400$ with
the larger lattice size for larger $\kappa$ and the number of time steps
to reach equilibrium were in the range $2\times 10^4 - 10^6$.
The distance between two pinning stripes, $d$, is given by the total length
of the solid in the $y$-direction, divided by the total number of stripes:
\begin{eqnarray}
d = {L \over c L/a}  & = & {a \over c} .
\label{d}
\end{eqnarray}
Therefore the critical density, $c^*_{\kappa}$, for which collective pinning
should
set in given by equating this distance $d$ with the capillary length $\kappa$
is:
\begin{eqnarray}
c^*_{\kappa}  & = & {a \over \kappa} .
\label{c*kappa}
\end{eqnarray}
For the given variables this gives the critical densities
$c^*_{\kappa} = 0.01 , 0.02, 0.04$.
Since $<h>$
increases linearly with $c$ for all the three values of $\kappa$  without
showing any sign of crossover, we conclude that we are in a single pinning
regime contrary to the simple argument
that we suggested above.

The reason for this discrepancy is that the pinning stripes introduce a
new length scale in the problem, namely the
distance between the height to which the liquid would rise without pinning
stripes $<h_0>$
minus the height to which the liquid would rise with a density of one of
pinning stripes, $<h_0'>$:
\begin{eqnarray}
H \equiv <h_0'> - <h_0> & = & \kappa (\sqrt{ (1- \sin (\theta )}-\sqrt{(1- \sin
(\theta')}) .
\label{H}
\end{eqnarray}
Therefore if $H$ plays a role for the onset of the single/collective pinning
regime
one should be able to go from one pinning regime to another
by changing {\em either} $|\theta' - \theta |$ {\em or}
$\kappa$.
This statement we have confirmed by keeping $\kappa$ constant and
increasing the quantity $|\theta' - \theta |$ for various values of $\kappa$
(see the discussion after  Fig.~3). Furthermore, in Fig.~2 is shown the
averaged height of the triple line $<h>$ versus the density $c$ for
different values of $\kappa$ for a fixed value of $|\theta' - \theta |$.
One notice a crossover from a single pinning regime to a collective pinning
regime as one increases $c$, and with the crossover appearing for smaller
$c$ the larger the value of $\kappa$.
For given values of $\kappa$, $\theta$, $\theta'$  and for small $c$'s a stripe
do not feel the presence of its neighbor stripes. As $c$ increases, the
curvature of the triple line between two stripes increases up to a point
where the curvature becomes so large, that the cost in gravitational energy
by lifting the liquid between two stripes is outbalanced by a decrease
in curvature and thereby surface energy. When this happens the system is in
the collective pinning regime.
Assuming that the decay of the triple line away from a stripe is exponential,
we estimate that curvature effects become important when the distance
the triple line has decayed after one correlation length, becomes on the order
of
the averaged distance between two stripes:
\begin{eqnarray}
{a \over c_{\rm curvature}^* } = d^*  & = & H exp(-1) \nonumber \\
c_{\rm curvature}^*  & = &
{a \over \kappa (\sqrt{ (1- \sin (\theta )}-\sqrt{(1- \sin (\theta')}) exp(-1)}
\label{c*curvature}
\end{eqnarray}
Assuming the form of the triple line in
between two pinning stripes can be described as a segment of a circle, another
way of stating Eq.~(\ref{c*curvature}) is to say that collective
pinning happens once the radius of curvature between the stripes becomes
on the order of the averaged distance between the stripes.
In Fig.~2 and Fig.~3 are indicated the critical densities
$c_{\rm curvature}^*$ for the various parameters used in the simulation.
Giving that Eq.~(\ref{c*curvature}) is only based on a simple order
of magnitude argument, the agreement with the onset of the collective
pinning deduced from the simulations is striking.
Besides, using Eq.~(\ref{c*curvature}) for the simulations performed in
Fig.~1, all give a $c_{\rm curvature}^*$ larger than one,  meaning that
collective pinning should not occur for these values of parameters,
which is in agreement with Fig.~1. Based on these results, we conjecture that
Eq.~(\ref{c*curvature}) should also give a good estimate in the case
of random pinning sites, with $\theta$ describing the averaged
value of the strengths of pinning sites, and $\theta'$ the fluctuations about
it.

\section{\bf Mean field theory }
We now propose a simple mean field theory in order to obtain the averaged
height
of the triple line, $< h_0(c,\kappa,a_0,\theta,\theta') >$, as a function of
the density
of pinning stripes $c$.
The mean field assumption amounts to consider just one pinning stripe,
and to find the minimum energy of this configuration. Furthermore we simplify
the
problem, and consider only a 2--dimensional projection of the 3--dimensional
liquid
meniscus onto the solid;
We will assume the functional form
for the height $h$ as a function of distance $y$ away from the pinning stripe:
\begin{eqnarray}
y \in [0:a] : h(y) = h_0 & ; &  y \in [a:d] : h(y) = h_0 \exp (-{y \over
\kappa}) + h_1
\label{hform}
\end{eqnarray}
The two constants $h_0$ and $h_1$ will be determined form the requirement that
the
mean field energy $E_{\rm mean}$ of the system attains it minimum. $E_{\rm
mean}$ is
given by:
\begin{eqnarray}
E_{\rm mean} & = & {1 \over 2} \rho g t h^2_0 a
- \gamma_{LG} \cos (\theta') h_0 a
- \int_{a}^{a+d} dy  \gamma_{LG} \cos (\theta) h_0 \exp (-{y \over \kappa})
-  \gamma_{LG} \cos (\theta) h_1 d \nonumber \\
& & + \int_{a}^{a+d} dy {1 \over 2} \rho g t h_0^2 \exp (-{2y \over \kappa})
+ \int_{a}^{a+d} dy {1 \over 2} \rho g t h_1^2
+ \int_{a}^{a+d} dy \rho g t h_0 h_1 \exp (-{y \over \kappa}).
\label{emean}
\end{eqnarray}
The variable $t$ is supposed to take into account the contribution in energy of
the
three dimensional meniscus on the two dimensional projection due to interfacial
and gravitational energy. The form of $t$ will be determined by the
requirement:
\begin{eqnarray}
< h_0(c,\kappa,a,\theta,\theta') > \rightarrow h_0  , c \rightarrow 0 & ; &
< h_0(c,\kappa,a,\theta,\theta') > \rightarrow h_1  , c \rightarrow 1
\label{condt}
\end{eqnarray}
Minimizing the energy
\begin{eqnarray}
& & {\partial E_{\rm mean} \over \partial h_0} = 0  ;
{\partial E_{\rm mean} \over \partial h_1} = 0 \nonumber \\
& & \rho g t h^2_0 a
- \gamma_{LG} \cos (\theta') a
(- \gamma_{LG} \cos (\theta) + \rho g t h_1 )
\kappa [\exp (-{a \over \kappa}) - exp (-{a+d \over \kappa}) ] \nonumber \\
& & + {1 \over 2} \rho g t h_0
\kappa [\exp (-{2a \over \kappa}) - exp (-{2(a+d) \over \kappa}) ] = 0 ;
\nonumber \\
& & - \gamma_{LG} \cos (\theta) d
+ \rho g t h_0 )
\kappa [\exp (-{a \over \kappa}) - exp (-{a+d \over \kappa}) ]
+ \rho g t h_1  d  = 0.
\label{h0h1}
\end{eqnarray}
Eq.~(\ref{h0h1}) is two equations with two unknowns from which $h_0, h_1$ can
be
determined. Now $< h_0(c,\kappa,a_0,\theta,\theta') >$ is given by the integral
over the functional form Eq.~(\ref{hform}):
\begin{eqnarray}
(a+d) < h_0(c,\kappa,a_0,\theta,\theta') > & =&
\int_{0}^{a} dx  h_0
+ \int_{a}^{a+d} dx [h_0 \exp (-{x \over \kappa}) + h_1],
\label{hform1}
\end{eqnarray}
which from the solutions of Eq.~(\ref{h0h1}) can be written:
\begin{eqnarray}
 < h_0(c,\kappa,a_0,\theta,\theta') > & = &
{\kappa^2 \gamma_{LG} \cos (\theta') a c \over
2 t [a - {\kappa^2 \over a(1/c -1)}
(\exp (-{a \over \kappa}) - exp (-{a \over c \kappa}))^2
+ {1\over 2} \kappa (\exp (-{2a \over \kappa}) - exp (-{2a \over c \kappa}) ]}
\nonumber \\
& & + {\gamma_{LG} \cos (\theta ) (1-c) \over \rho g t}
\label{hform2}
\end{eqnarray}
In order to fulfill the condition Eq.~(\ref{condt}) it can be seen from
Eq.~(\ref{hform2})
that a proper choice of $t$ is
\begin{eqnarray}
t & =&
\sqrt{{ \gamma_{LG} \over 2 \rho g [c(1-\sin (\theta') + (1-c)(1-\sin(\theta )
]}}
(c\cos (\theta' ) + (1-c)\cos (\theta ) )
\label{tform}
\end{eqnarray}
Thus we end up with the form for $<h>$ given by:
\begin{eqnarray}
 < h_0(c,\kappa,a_0,\theta,\theta') > & =&
{\kappa^2 \gamma_{LG} \cos (\theta') a c \over
2 t [a - {\kappa^2 \over a(1/c -1)}
(\exp (-{a \over \kappa}) - exp (-{a \over c \kappa}))^2
+ {1\over 2} \kappa (\exp (-{2a \over \kappa}) - exp (-{2a \over c \kappa}) ]}
\nonumber \\
& & + {\gamma_{LG} \cos (\theta ) (1-c) \over \rho g t}
\label{hform3}
\end{eqnarray}
Finally we will assume that the width of the pinning stripes $a$ depends on the
density of pinning stripes $c$ in the following way:
\begin{eqnarray}
c \in [0:0.5] : a = a_0 & ; &  c \in [0.5:1] : a = a_0 L 2(c-0.5)
\label{aversusc}
\end{eqnarray}
which just states that the no two pinning stripes will be neighbors until
$c=0.5$ where after
the width of a stripe increases linearally with $c$.

In Fig.~3 is shown
the averaged height of the triple line $<h>$ versus the density $c$ for
different values of $\theta$ with a fixed value of $\kappa$ and $\theta'$.
One notice that the density for which collective pinning begins clearly depends
on $\theta$ and is well described by Eq.~(\ref{c*curvature}). Furthermore the
2-dimensional mean field solution gives a good description of the 3-dimensional
Langevin solution, with the best agreement for small $H$ in  Eq.~(\ref{H}).
This
is to be expected since for large $H$ the 3-dimensional nature of the liquid
meniscus becomes    more important and the 2-dimensional mean field picture
breaks down.

In conclusion we have proposed a
macroscopic Hamiltonian approach to the pinning of solid--liquid--vapour
interfaces due to presence of stripes of obstacles. We find that curvature
effects play a crucial role for the transition from the single pinning regime
to the collective pinning regime. We have proposed an estimate of the density
of pinning stripes for which the collective pinning happens, which is in
good agreement with the simulations of the Langevin equation. We conjecture
the same estimate to be valid in the case of random pinning sites. Finally a
2-dimensional mean field solution has been introduced which for small values
of $|\theta' - \theta |, \kappa$ gives excellent approximation for the height
of the
triple line.

J.V.A. wish to acknowledge support from
the Danish Natural Science Research Council under Grant No. 9400320.

\begin{figure}
\begin{center}
\caption{The averaged height of the triple line $<h>$ versus the density $c$
for
different values of $\theta$.
$\kappa = 0.5, \theta' = 1.0 $.
$\theta = 0.5 (\times)$, $\theta = 1.2 (\bigtriangleup)$,
$\theta = {\pi \over 2} (\Diamond)$,
$\theta = -0.5 (+)$ and $\theta = -1.0 (\Box)$ respectively. The
solid lines are obtained from the mean field solution Eq.~(\ref{hform3})
{}From Eq.~(\ref{c*curvature}) one finds:
$c_{\rm curvature}(\theta =0.5)=0.67,
c_{\rm curvature}(\theta =1.2)>1,
c_{\rm curvature}(\theta = {\pi \over 2})=0.55,
c_{\rm curvature}(\theta =-1.0)=0.23$, and
$c_{\rm curvature}(\theta =-0.5)=0.27$.}
\end{center}
\end{figure}

%
%

\begin{thebibliography}{99}
\bibitem{note2}
After publication of \cite{ego} we learned that a similar approach
like Eq.~(\ref{hamiltonian}) was used in \cite{Schwarts}, but with double
periodic wettability in $y$ and $z$ with wavelengths $\lambda$, $lambda_{\nu}$
respectively. Despite the similarity of the Hamiltonian
Eq.~(\ref{hamiltonian}),
the analysis in \cite{Schwarts} was quite different from \cite{ego} and what
follows in this paper.
\bibitem{note1}
$T$ is the temperature as a function of Monte Carlo time $t$; in \cite{ego}
$T$ was chosen proportional to $t$, i.e. $T(t) = T_0 (1 - {t \over N})$ with
$N$ the total number of Monte Carlo steps and $T_0$ the initial temperature.
\end{thebibliography}
\end{document}